# Far-Field Microscopy of Sparse Subwavelength Objects

**A. Szameit[1], Y. Shechtman[1], H. Dana[2], S. Steiner[3], S. Gazit[1], T. Cohen-Hyams[4], E. Bullkich[1], O. Cohen[1], Y. C. Eldar[5], S. Shoham[2], E. B. Kley[3], and M. Segev[1]**

[1]*Physics Department and Solid State Institute, Technion, 32000 Haifa, Israel*

[2]*Department of Biomedical Engineering, Technion, 32000 Haifa, Israel*

[3]*Institute of Applied Physics, Friedrich Schiller University Jena, Max-Wien-Platz 1, 07743 Jena, Germany*

[4]*Microelectronics Research Center, Department of Electrical Engineering, Technion, 32000 Haifa, Israel*

[5]*Department of Electrical Engineering, Technion, 32000 Haifa, Israel*

**Abstract**

We present the experimental reconstruction of sub-wavelength features from the far-field of sparse optical objects. We show that it is sufficient to know that the object is sparse, and only that, and recover *100 nm* features with the resolution of *30 nm*, for an illuminating wavelength of $\lambda=532\ nm$. Our technique works in real-time, requires no scanning, and can be implemented in all existing microscopes - optical and non-optical.

## Introduction and Background

The wavelength $\lambda$ is the fundamental scale in the theory of imaging since the days of Ernst Abbe, when it was shown that the maximal resolution of a perfect optical imaging system arises from the numerical aperture of the lenses involved, with the best recoverable resolution of $\lambda/2$ [1]. This concept is deeply rooted in the fundamental principles of electromagnetic (EM) wave propagation. The transfer function for EM waves in free space renders spatial frequencies larger than $1/\lambda$ evanescent [2], decaying rapidly on a distance of several wavelengths. Hence, in the optical far-field their amplitude is buried in the noise (from spurious scattering and from the detection system). It is generally agreed that, using conventional imaging methods, the sub-wavelength features of the propagating light are irretrievably lost. Throughout the years, there were many attempts to go beyond the $\lambda/2$ resolution limit, either by scanning at very close proximity ("near field") to the sub-wavelength specimen [3-5], by scanning with a sub-wavelength super-oscillatory hot-spot [6], by using negative-index materials [7-12], or by placing sub-wavelength fluorescing items on the object [13,14]. However, all of these methods have shortcomings: near-field methods require placing the detection device in the near field of the specimen and scanning the sample point-by-point, thus preventing real-time operation, and methods relying on negative-index materials involve heavy losses with current technology. Likewise, techniques relying on random distribution of fluorescent molecules require averaging over multiple experiments – rendering real-time imaging impractical, and of course the fluorescing items affect the chemical processes in their vicinity, which is generally undesirable when studying real-time activity of biological objects, such as living cells, bacteria, etc. Over the years, various algorithmic approaches were also suggested, utilizing the analytic character of electromagnetic waves, which allows analytic extrapolation of the optical field in all three spatial dimensions (see, e.g. [15,16]). However, all of these methods greatly suffer from noise in the measured data, and are generally considered impractical for imaging at sub-wavelength resolution [17]. Recently, we demonstrated a new method for reconstructing sub-wavelength features from the far-field of sparse samples [18,19], based on ideas of Compressed Sensing (CS) [20-24]. Our technique is robust to noise, and relies only on the advance knowledge that the object is sparse in a known basis, and nothing more [25]. Here, for the first time, we present the experimental realization of sub-wavelength imaging: we recover *100 nm* features at a *30 nm* resolution, using *532 nm* light. In doing that, we reconstruct features that are 3.5 times smaller than diffraction limit, at a resolution 8 times better than this limit.

It is an essential feature of propagating light that, even in free space, only spatial frequencies below $1/\lambda$ can propagate, whereas all frequencies above $1/\lambda$ are rendered evanescent and decay exponentially. Hence, evanescent waves are buried in the noise after propagating a distance of merely a few wavelengths. Therefore, optical objects with sub-wavelength features appear highly blurred in a microscope, due to the loss of their high spatial-frequencies. Consequently, the goal of subwavelength microscopy can be translated into the reconstruction of the entire spatial frequency range of the information ("object"), by measuring solely frequencies in the range $[-1/\lambda, 1/\lambda]$. However, in mathematical terms, this bandwidth extrapolation problem corresponds to an under-determined system of equations, which cannot be inverted. The problem arises from the fact that such a system has an infinite number of solutions, which all produce the same (blurred) image as seen in the microscope. That is, using the measured data carried by the propagating spatial frequencies (either by measuring the far field or by measuring the blurred image), one can add any information in the evanescent part of the spectrum, and the measured data will remain unchanged, while of course only one choice corresponds to the desired sub-wavelength information that was cut off by diffraction limit. The crucial task is therefore to extract the one correct solution, which is, in other terms, the solution “that makes the most sense”. This is where sparsity comes into play. When decomposing a light field into a set of basis functions, this field is said to be sparse when most of the projections of the basis functions are negligibly small. For example, an optical object is sparse in the near-field when the number of non-zero pixels is small compared to the entire field of view. In our recent paper [18], we have demonstrated theoretically a new method for reconstructing sub-wavelength features from the far-field of sparse samples. Our technique relies only on the advance knowledge that the optical object is sparse in a known basis (e.g., in real-space), and nothing more. We have given an experimental proof of concept for super-resolution, where we were able to reconstruct features that were many times smaller than the finest feature defined by the numerical aperture of the system [18,19]. Here, we take this idea into the sub-wavelength domain, and demonstrate the construction of *100 nm* features with the resolution of *30 nm*, for an illuminating wavelength of $\lambda=532$ *nm*. Our technique works in real-time, requires no scanning, and can be implemented in all existing microscope, optical and non-optical.

**Intuition & Explanation**

As explained above, sub-wavelength imaging can be represented as a bandwidth extrapolation problem, where propagation is free space is equivalent to going through a band-limited transfer function, with the diffraction limit acting as the cutoff frequency of a low-pass filter [17]. In principle, the bandwidth can be extrapolated by any function, and the measured (low-pass filtered) data would not change. The question is how to identify the correct extrapolation, out of the infinite number of possibilities. Here comes sparsity into play. It is the essential result of CS that, in the absence of noise, if the information (to be recovered) is sparse in a basis that is sufficiently uncorrelated with the measurement basis, then the sparsest solution is unique. For our particular case of sub-wavelength imaging, there is only one continuation of this truncated spatial spectrum that yields the sparsest image. That is, from all the possible solutions which could create the truncated far-field seen in the microscope, only one is sparse. Hence, if we ***know*** that our image (information to be recovered) is sparse in real-space, and only that, we just need to find the only sparse solution which generates the observed far-field image. The uniqueness of the solution guarantees that this is the correct one. In the presence of noise, we find empirically (in simulations) that the solutions, although not unique anymore, are all nevertheless very close to the original information when the noise is small enough. Hence, searching for the sparsest solution (under the constraint that the measured data would not change), yields a reconstruction that is very close to the ideal one under typical experimental conditions. Finding the sparsest solution can be done through various algorithms. For electromagnetic fields, which are complex functions (because they generally carry non-uniform phase), we have demonstrated the nonlocal hard thresholding algorithm [18], which can reconstruct both amplitude and phase. In the experiments presented below, the optical information we wish to recover carries no phase (we use a set of holes in a planar sample), hence we use the Basis Pursuit algorithm.

**Experiments**

We now demonstrate our technique of sub-wavelength imaging experimentally, on both one-dimensional (1D) and two-dimensional (2D) structures. For sub-wavelength imaging of a 1D structure, we fabricate a sparse sample consisting of only four narrow stripes, each $20\ \mu m$ long and *150 nm* wide [26]. The four stripes are arranged in two pairs; within each pair the stripes are spaced by *150 nm*, while the pairs are separated by *300 nm*. For further information on the samples, see the Methods Section. Figure 1a shows a Scanning Electron Microscope (SEM) image of the sample. When illuminating such a sample with monochromatic light at $\lambda=532$ nm, this sample practically represents a one-dimensional sub-wavelength problem, and is therefore well suited to demonstrate our reconstruction technique. In our hand-made microscope with a numerical aperture close to 1, we observe a small and blurred image, where the two stripes within each pair merge and are indistinguishable, as shown in Fig. 1b. When imaging the optical far-field (Fig. 1c) it is evident that it covers much more pixels than the blurred image, therefore facilitating a much higher number of meaningful measurements (since each pixel corresponds to one measurement). Searching for the sparsest solution (with the mathematical techniques described in the Methods Section) yields the reconstruction presented in Fig. 1d, which closely resembles the SEM image of our sample (Fig. 1a). Figure 1e compares the cross sections of the original sample and the reconstructed image. The reconstruction coincides with the original information in the widths and positions of the stripes, including the separation between stripes. There are small discrepancies in the stripes heights, and in the little additional pedestals accompanying the reconstructed stripes. These small discrepancies arise from systematic errors in our hand-made microscope (e.g., small tilt in the measured wavefront, etc.), and should vanish for a perfectly calibrated system. That is, we envision placing the sample on a microscope slide, within a "useful region" of, say, 10-30μm, around which there are known features of above-wavelength size – which can be used for calibrating the alignment of the far-field plane. The performance or our approach is even more evident in the spatial spectrum (Fig. 1e). The black dots represent the measurements, which stop at the cutoff at $1/\lambda$. The red solid line shows the reconstructed spatial frequencies, up to $8/\lambda$. For comparison, the black dashed line depicts the spatial frequencies of the original sample (calculated by Fourier transforming the SEM image). Clearly, the reconstructed power spectrum almost fully coincides with the true spatial spectrum, demonstrating an increase of the resolution by a factor of 8. In numbers, at $\lambda=532\ nm$ this corresponds to a resolution of $\approx 30\ nm$.

Our technique works very well also in two dimensions. To this end, we fabricated a Star of David, consisting of 30 sub-wavelength holes, with a diameter of *100 nm* each, spaced by *100 nm*. Figure 2a shows an SEM image of this sample. In our microscope, the observed image is very small and highly blurred, as the one dimensional structure of Fig. 2b. However, the far-field of the sample, obtained by removing the imaging lens, contains much more details with high resolution, as shown in Fig. 2c. Taking the data from this image and searching for the sparsest solution reproducing it, we reconstruct a Star of David, as presented in Fig. 2d. The accuracy is good, although the reconstruction still has some discrepancies in the positions of some of the holes. These arise again from systematic errors in our hand-made microscope, which we expect to disappear in a perfectly calibrated system. However, despite these small inaccuracies, one can safely state that the resolution is again around *30 nm*, at an illumination wavelength of $\lambda=532\ nm$.

**Discussion and Conclusions**

In this work, we have presented a technique facilitating real-time reconstruction of sub-wavelength features, at an unprecedented resolution for single-shot experiments. The method relies on prior knowledge – that the sample is sparse in a known basis (spatial near-field, in the examples here), and only that. It is hereby important to note that most natural and artificial objects are sparse, in some basis. The information does not necessarily have to be sparse in real space: it can be sparse in any mathematical basis whose relation to the measurement basis is known, e.g. the wavelet basis or the second derivative of the sample's profile. In all of these cases, our technique can provide a major improvement by "looking beyond the resolution limit" in a single-shot experiment. Since our approach is purely algorithmic, it can be applied to every optical microscope as a simple computerized image processing tool, delivering results in real time with practically no additional hardware. Our technique is very general, and can be extended also to other, non-optical, microscopes, such as atomic force microscope, scanning-tunnelling microscope, magnetic microscopes, and other imaging systems. We believe that the microscopy technique presented here holds the promise to revolutionize the world of microscopy with just minor adjustments to current technology: sparse sub-wavelength images could be recovered by making efficient use of their available degrees of freedom. Finally, we would like to emphasize that our approach is much more general than the particular subject of optical sub-wavelength imaging. It is in fact a universal scheme for recovering information beyond the cut-off of the response function of a general system, relying only on priori information that the information is sparse in a known basis. As an exciting example, we have recently investigated the ability to utilize this method for recovering the actual shape of very short optical pulses measured by a slow detector. Our preliminary theoretical and experimental results indicate, unequivocally, that our method offers an improvement by orders of magnitude beyond the most sophisticated deconvolution methods. In a similar vein, we believe that our method can be applied for spectral analysis, offering a means to recover the fine details of atomic lines, as long as they are sparse (i.e., do not form bands). In principle, the ideas described here can be generalized to any sensing

/ detection / data acquisition schemes, provided only that the information is sparse in a known basis, and nothing more.

## Methods

A – Searching for the sparsest solution

Finding the sparsest object (in a known basis) that creates the measured far-field can be formulated as the mathematical problem

$$(P_0): \min_x \|x\|_0 \ \textit{subject to} \ \|FAx - b\|_2 < \varepsilon ,$$

where $x$ is a sparse vector representing the (only few non-zero) coefficients of the unknown object, in a known mathematical basis represented by the matrix $A$. The matrix $F$ represents the transfer function of the imaging system, $b$ is a vector representing the measured image, and $\varepsilon$ is a parameter determined by the measurement noise. In the particular case presented in this paper, $x$ is in real-space and b, the measurement vector, is in the far field (Fourier space). The term $\|x\|_0$ denotes the $L_0$ norm, which simply counts the number of non-zero elements in the vector $x$. It can be shown that ($P_0$) is a problem that can be generally solved only in non-polynomial time [27]. However, there exist approximate solutions to this problem, in particular the convex approximation

$$(P_1): \min_x \|x\|_1 \ \textit{subject to} \ \|FAx - b\|_2 < \varepsilon$$

where the $L_0$ norm is replaced by an $L_1$ norm. It has been rigorously proven that the solutions of ($P_1$) are close to that of ($P_0$), given that $x$ is sufficiently sparse [28]. This convex approximation is known as Basis Pursuit [29], and can be solved with various standard methods: Searching among all solutions $x$ that solve $\|FAx - b\| < \varepsilon$ for the one with the smallest $L_1$ norm. This is the methodology we use here. More details on the mathematical formulation can be found in [18]. We note that, as we have shown in [18], the concept of sparsity-based super-resolution and sub-wavelength imaging works also when the measurements are taken in the image plane, that is, $x$ and $b$ can both be in the same (known) basis.

B – Fabrication of the samples generating the optical information

The optical information is generated by passing a collimated laser beam through a mask, whose transmission function corresponds to the optical information superimposed on the laser beam. The mask is fabricated as follows: As substrate material we chose fused silica, because it is a high quality transparent material at optical frequencies, and because its processing technology is well developed. In order to to create a mask containing the optical image, we deposit some opaque material on the substrate and make some patterned holes in it, such that the holes pass the light while the opaque material blocks it. For this purpose, we sputter a chromium layer onto the surface of the substrate. Chromium is a metal, which absorbs light at optical frequencies. Nevertheless, the thickness of the chromium layer has to be larger than the skin depth at optical frequencies, to avoid unintended transmission through that layer. Thus we select a thickness of *100 nm* as suitable compromise between high quality optical behavior and fabrication considerations. The structures in the chromium layer are nano-holes, drilled in the chromium by a beam of focused gallium ions from a liquid metal ion source [30, 31] (Zeiss Neon 60). With this technology, it is feasible to mill the desired structures into the chromium layer directly and efficient, without any additional lithography process. Utilizing a convenient set of parameters, it is possible to imprint the designed structures into the metal layer, without significantly affecting the substrate material, and with high spatial accuracy (cf. SEM images).

We fabricated two different samples: (1) four narrow stripes, each *20 μm* long and *150 nm* wide, which are arranged in two pairs (separated *300 nm* from each other), which, when illuminated with 532nm light, yield a one-dimensional sub-wavelength optical structure, and (2) a Star of David, consisting of 30 holes, with *100 nm* diameter each, spaced by *100 nm* yielding a two-dimensional sub-wavelength optical structure.

## References

[1] E. Hecht. *Hecht Optics*. Addison-Wesley, 1998.

[2] M. Teich and B. Saleh. *Fundamentals of Photonics*. Wiley, New York, 1991.

[3] E.A. Ash and G. Nicholls. Super-resolution Aperture Scanning Microscope. *Nature*, 237(5357):510–512, 1972.

[4] A. Lewis, M. Isaacson, A. Harotunian, and A. Muray. Development of a 500-Å Spatial-Resolution Light-Microscope: I. Light is Efficiently Transmitted Through λ/16 Diameter Apertures. *Ultramicroscopy*, 13:227, 1984.

**[5]** E. Betzig, J.K. Trautman, T.D. Harris, J.S. Weiner, and R.L. Kostelak. Breaking the diffraction barrier: optical microscopy on a nanometric scale. *Science*, 251(5000):1468–1470, 1991.

[6] F. M. Huang and N. I. Zheludev, "Super-resolution without evanescent waves," Nano Lett. **9**, 1249–1254 (2009).

[7] J. B. Pendry, "Negative refraction makes a perfect lens," Phys. Rev. Lett. **85**, 3966–3969 (2000).

[8] N. Fang, H. Lee, C. Sun, and X. Zhang, "Sub-diffraction-limited optical imaging with a silver superlens," Science **308**, 534–537 (2005).

[9] Z. Jacob, L. V. Alexeyev, and E. Narimanov, "Optical hyperlens: far-field imaging beyond the diffraction limit," Opt. Express **14**, 8247–8256 (2006).

[10] A. Salandrino and N. Engheta, "Far-field subdiffraction optical microscopy using metamaterial crystals: Theory and simulations," Phys. Rev. B **74**, 075103 (2006).

[11] Z. Liu, H. Lee, Y. Xiong, C. Sun, and X. Zhang, "Far-field optical hyperlens magnifying sub-diffraction-limited objects," Science **315**, 1686 (2007).

[12] I. I. Smolyaninov, Y. J. Hung, and C. C. Davis, "Magnifying superlens in the visible frequency range," Science **315**, 1699–1701 (2007).

[13] A. Yildiz, J. N. Forkey, S. A. McKinney, T. Ha, Y. E. Goldman, and P. R. Selvin, "Myosin v walks hand-overhand: Single fluorophore imaging with 1.5nm localization," Science **300**, 2061–2065 (2003).

[14] S.W. Hell, R. Schmidt, and A. Egner, "Diffraction-unlimited three-dimensional optical nanoscopy with opposing lenses," Nat. Photon. **3**, 381–387 (2009).

[15] A. Papoulis, “A new algorithm in spectral analysis and band-limited extrapolation,” IEEE Trans. Circuits Syst. **22**, 735–742 (1975).

[16] R. W. Gerchberg, “Super-resolution through error energy reduction,” J. Mod. Opt. **21**, 709–720 (1974).

[17] J. W. Goodman, *Introduction to Fourier optics* (Englewood, CO: Roberts & Co. Publishers, 2005), 3rd ed.

[18] S. Gazit, A. Szameit, Y. C. Eldar, and M. Segev, “Reconstruction of sparse subwavelength images,” Opt. Exp. **26**, 23920-23946 (2009).

[19] Y. Shechtman, S. Gazit, A. Szameit, Y. C. Eldar, and M. Segev, “Super-resolution and reconstruction of sparse images carried by incoherent light,” Opt. Lett. **35**, 1148-1150 (2010).

[20] E. J. Candes, J. Romberg, and T. Tao, “Robust uncertainty principles: exact signal reconstruction from highly incomplete frequency information,” IEEE Trans. Inf. Theory **52**, 489–509 (2006).

[21] E. J. Candes and T. Tao, “Near-optimal signal recovery from random projections: Universal encoding strategies?” IEEE Trans. Inf. Theory **52**, 5406–5425 (2006).

[22] E. J. Candes and M. B.Wakin, “An introduction to compressive sampling,” IEEE Signal Process. Mag. **25**, 21–30 (2008).

[23] D. L. Donoho, “Compressed sensing,” IEEE Trans. Inf. Theory **52**, 1289–1306 (2006).

[24] Note that, in the past two years, CS was utilized also in optics for reducing the sampling rate in a single-pixel camera [M. F. Duarte *et al.*, IEEE Sig. Proc. Mag. **25**, 83–91 (2008)] and in ghost imaging [O. Katz, Y. Bromberg, and Y. Silberberg, Appl. Phys. Lett. **95**, 131110 (2009)].

[25] Last month, another group has proposed theoretically using CS for super-resolution imaging [Y. Rivenson, A. Stern, and B. Javidi, Opt. Exp. **18**, 15094–15103 (2010)], in a fashion that is conceptually similar to ours.

[26] The sample is made of *100 nm* thick chromium layer on glass, and the optical image is in fact holes made in the chromium. The thickness of the chromium layer is larger than the skin depth at optical frequencies.

[27] A. M. Bruckstein, D. L. Donoho, and M. Elad, “From Sparse Solutions of Systems of Equations to Sparse Modeling of Signals and Images,” SIAM Rev. **51**, 34-81 (2009).

[28] D.L. Donoho and X. Huo, “Uncertainty Principles and Ideal Atomic Decomposition,” IEEE Trans. Inf. Theory **47**, 2845-2862 (2001).

[29] S.S. Chen, D.L. Donoho, and M. A. Saunders, "Atomic decomposition by basis pursuit," SIAM review **43**, 129-159 (2001).

[30] V. Krohn and G. Ringo, "Ion source of high brightness using liquid metal" Appl. Phys. Lett. **27**, 479 (1975).

[31] P.D. Prewett and D.K. Jefferies, "Characteristics of a Gallium Liquid Metal Field Emission Ion Source", J. Phys. D **13**, 1747 (1980).

## Figure captions

**Caption Fig. 1:**

**Reconstruction of one-dimensional sub-wavelength information**

(a) SEM image of the sample. Since the stripes are 20 μm long, at an illumination wavelength of $\lambda$=*532 nm*, this sample practically represents a one-dimensional sub-wavelength problem. (b) The blurred image of the sample, as seen in the microscope. The four stripes are merged into two. (c) The optical far-field of the 1D information, observed by removing the imaging lens of the microscope. (d) The reconstructed 1D information, recovered from the measured far-field, using only the fact that the information is sparse in real-space. (e) Comparison of the 1D reconstructed information and the cross-section of the original image. (f) The recovered spatial frequencies extend to almost $8/\lambda$, when only spatial frequencies below $1/\lambda$ are measured. This corresponds to a resolution of *30 nm*, at $\lambda$=*532 nm* illumination wavelength.

**Caption Fig. 2:**

**Reconstruction of two-dimensional sub-wavelength information**

(a) SEM image of the sample. (b) The blurred image of the sample, as seen in the microscope. The individual holes cannot be resolved. (c) The optical far-field of the sample, observed when removing the imaging lens of the microscope. (d) The reconstructed 2D information, recovered from the measured far-field, using only the fact that the sample is sparse in real-space.

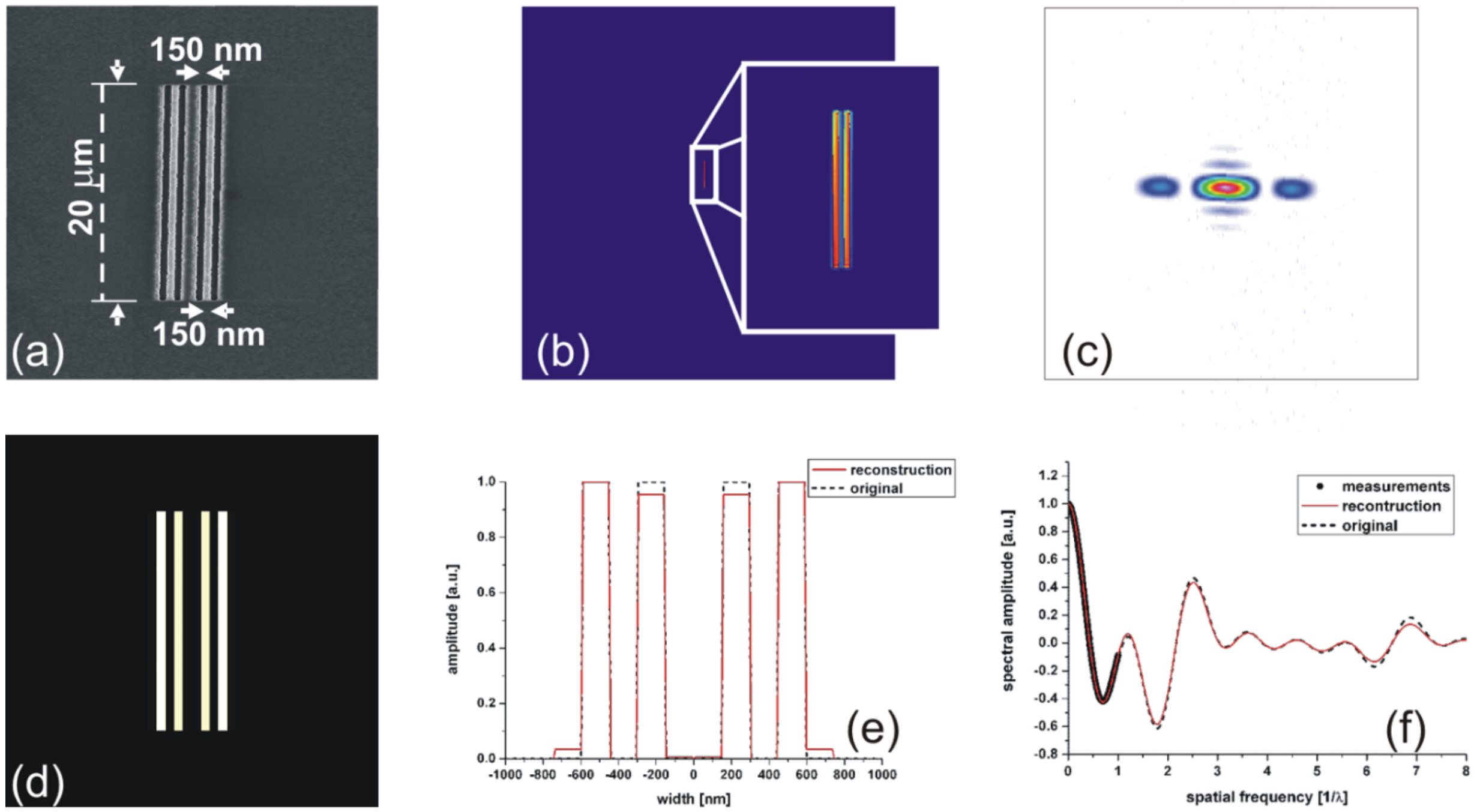
150 nm
20 μm
150 nm
(a)
(b)
(c)
(d)
reconstruction
original
amplitude [a.u.]
width [nm]
(e)
measurements
recontruction
original
spectral amplitude [a.u.]
spatial frequency [1/λ]
(f)

**Fig. 1**

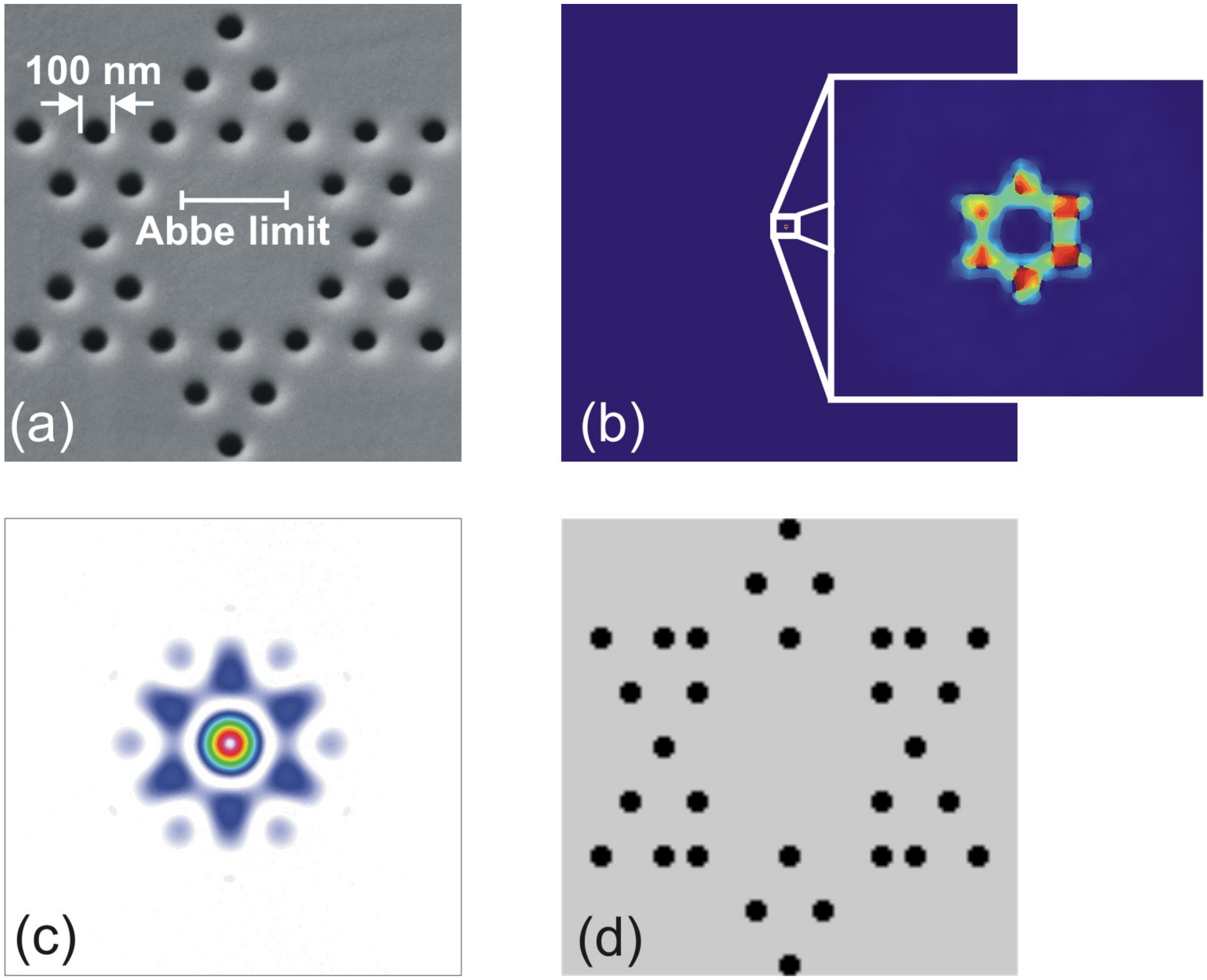

**Fig. 2**